\documentclass[useAMS,usenatbib]{mn2e}

\usepackage{times,graphicx,amssymb,natbib}

\title[Stellar Encounters: A Stimulus for Disc Fragmentation?]{Stellar Encounters: A Stimulus for Disc Fragmentation?}
\author[Duncan Forgan and Ken Rice]{Duncan Forgan $^{1}$\thanks{E-mail:
dhf@roe.ac.uk} and Ken Rice$^{1}$\\
$^{1}$Scottish Universities Physics Alliance (SUPA), Institute for Astronomy, University of Edinburgh, Blackford Hill, Edinburgh, EH9 3HJ, Scotland, UK \\
}
\begin{document}

\date{Accepted 0000}

\pagerange{\pageref{firstpage}--\pageref{lastpage}} \pubyear{0000}

\maketitle

\label{firstpage}

\begin{abstract}

\noindent An interaction between a star-disc system and another star will perturb the disc, possibly resulting in a significant modification of the disc structure and its properties.  It is still unclear if such an encounter can trigger fragmentation of the disc to form brown dwarfs or gas giant planets.  This paper details high resolution Smoothed Particle Hydrodynamics (SPH) simulations investigating the influence of stellar encounters on disc dynamics.  Star-star encounters (where the primary has a self-gravitating, marginally stable protostellar disc, and the secondary has no disc) were simulated with various orbital parameters to investigate the resulting disc structure and dynamics.  This work is the first of its kind to incorporate realistic radiative transfer techniques to realistically model the resulting thermodynamics. 

The results suggest that the effect of stellar encounters is to prohibit fragmentation - compressive and shock heating stabilises the disc, and the radiative cooling is insufficient to trigger gravitational instability.  The encounter strips the outer regions of the disc (either through tidal tails or by capture of matter to form a disc around the secondary), which triggers a readjustment of the primary disc to a steeper surface density profile (and a flatter Toomre Q profile).  The disc around the secondary plays a role in the potential capture of the secondary to form a binary.  However, this applies only to orbits that are parabolic - hyperbolic encounters do not form a secondary disc, and are not captured.

\end{abstract}


\section{Introduction}

\noindent Disc fragmentation has been touted as a possible process for forming brown dwarfs and gas giant planets \citep{Boss_science,Stam_frag}.  If the conditions are right, it does indeed seem possible that bound objects can be formed in this top-down process: what is less clear is if these conditions exist in nature.  There are two criteria that must be satisfied before fragmentation can occur, both of which relate to cooling processes \citep{Gammie,Ken_1,Durisen_review}. The first condition that the disc must satisfy is that it must be (at least locally) gravitationally unstable under perturbation, to allow fragmentation to begin.  A disc is  gravitationally unstable to axisymmetric perturbations if the value of the Toomre parameter \citep{Toomre_1964}

\begin{equation} Q = \frac{c_s \kappa}{\pi G \Sigma} \sim 1, \label{eq:Q}\end{equation} 

\noindent where \(c_s\) is the sound speed, \(\kappa\) is the epicyclic frequency (which is equal to the angular frequency \(\Omega\) if the disc is Keplerian), and \(\Sigma\) is the disc surface density.  Nonaxisymmetric perturbations require \(Q<1.5-1.7\) for instability (see \citealt{Durisen_review}).  If this condition is satisfied, perturbations can then grow on the local dynamical timescale. 

The second condition refers to the growth of these perturbations.  As a perturbation grows, it will undergo heating due to compression and shocks. Therefore, if any perturbation is to continue to grow, it must be able to radiate this energy away efficiently, such that the cooling time is comparable to the local dynamic timescale:

\begin{equation} t_{cool} \leq  3\Omega^{-1} \end{equation}

\noindent \citep{Gammie,Ken_1,Mejia_2}.  If this condition is not met, the fragment cannot collapse to a sufficiently bound state to survive in the disc (typically, these unbound fragments are destroyed by Keplerian shear).  Again, cooling processes play an important role: if cooling is not efficient enough, the gas will be heated by compression and shocks, \(c_s\) will increase, and \(Q\) will therefore increase to above the critical value, stabilising the region.  Further, most discs will evolve until they reach their minimum \(Q\) value (representing an equilibrium between heating and cooling in the disc).  Once a disc reaches this state, it is said to be \emph{marginally gravitationally stable} \citep{Lodato_Rice_04}.

Recent work \citep{Matzner_Levin_05,Rafikov_05,Whit_Stam_06,Rafikov_07} suggests that the cooling processes in protoplanetary discs are of limited efficacy (which increases the minimum \(Q\) value).  Recent semi-analytical studies of angular momentum transport in discs by \citet{Rice_and_Armitage_09} and \citet{Clarke_09} show that the inner regions of discs are stable against fragmentation, and that gaseous material must orbit at distances of \emph{at least} 20 au (and possibly as distant as 70 au) if it is to become susceptible to fragmentation.   This confirms the work of numerical simulations \citep{Stam_frag} and is consistent with observations \citep{Greaves_Tau} that indicate disc fragmentation does not occur in the optically thick inner regions of discs, and will only fragment in the outer regions of discs with particular properties, e.g. discs that are massive and extended.  However, there is another way to decrease \(Q\): increase \(\Sigma\).  All the variables in equation (\ref{eq:Q}) (with the obvious exception of G) are local: hence it is possible to make the disc locally unstable by increasing the local density.  This is in principle achievable by the presence of another star moving through (or near) the disc.  The presence of a companion in the star-disc system can draw out the disc matter into tidal tails, compressing the gas while reducing its optical depth, improving its cooling efficiency.  Also, angular momentum transfer between the secondary and the disc can trigger outward angular momentum transport - this results in inward mass transport, increasing density in the inner regions.  If this forced increase in \(\Sigma\) by an encounter (which can help satisfy condition 1) can be accompanied by sufficient cooling (to satisfy condition 2), then in principle it is possible that \(Q\) can be maintained at a low value, and the disc can successfully fragment.  Again, this illustrates the importance of correctly treating radiative cooling in order to correctly ascertain whether disc fragmentation by this effect is feasible.

Some previous simulations suggest encounters can destabilise discs, and can cool sufficiently to cause fragmentation (e.g. \citealt{Boffin,Watkins_a,Watkins_b,Lin_et_al_98}), whereas others suggest encounters stabilise discs by compressive and shock heating, and prohibit fragmentation (e.g. \citealt{Lodato_encounters}). There have been many attempts to simulate such encounters: until recently, most analyses of the problem have not considered the full effects of radiative transfer, or have used simple cooling time parametrisations \citep{Lodato_encounters}.  The correct treatment of radiative transfer (to correctly evaluate the local cooling time) is crucial: if the disc is compressed by a companion, this compression should occur on the dynamical timescale, which is shorter than the thermal timescale.  Therefore, it is to be expected that the disc's subsequent evolution will be adiabatic, and hence fragmentation will not occur unless the cooling time can be reduced significantly by the compression, i.e.

\begin{equation} \left(\frac{dt_{cool}}{d\Sigma}\right)_S < 0 \end{equation}

\noindent This may be possible if the subsequent temperature increase pushes the gas into the so-called ``opacity gap" (where ice and dust evaporate), but the decrease in opacity reduces in magnitude as the density of the gas is increased, so fragmentation still appears to be unlikely \citep{Johnson_and_Gammie_03}.

This paper presents the results of a series of high resolution Smoothed Particle Hydrodynamics (SPH) simulations with radiative transfer \citep{intro_hybrid} of the encounter of a star-disc system with a discless companion.  The parameter space explored by these simulations has been selected to emulate the work of \citet{Lodato_encounters}, as a means of comparing the differences between using a cooling time formalism and more realistic approximations to radiative transfer.  By doing so, the effect of changing various orbital properties can be investigated.  In summary: \emph{no simulations carried out in this work exhibit fragmentation}.  Although the encounter does increase \(\Sigma\), radiative cooling is insufficient to overcome the significant compressive and shock heating.  The paper is organised as follows: section \ref{sec:method} describes the configuration of the numerical simulations performed; section \ref{sec:results} describes the results of the simulations, comparing the effects of changing the various orbital parameters; in sections \ref{sec:discussion} \& \ref{sec:conclusions} the results are summarised, and the paper is concluded.

\section{Method} \label{sec:method}

\begin{table*}
\centering
  \caption{Summary of the orbital parameters investigated in this work.\label{tab:params}}
  \begin{tabular}{c || cccccccc}
  \hline
  \hline
   Simulation  &  $M_{disc}/M_{\odot}$   & $\Sigma \propto r^{-x}$  &  $M_{2}/M_{\odot}$   & Calculated $R_{peri}$ (au)  & Actual $R_{peri}$ (au)& $e$ &   Prograde/Retrograde &   Inclination    \\  
 \hline
  1 & 0.1 & 1 & 0.1 & 40 & 28 & 1 & Pro & $0^{\circ}$ \\
  2 & 0.1 & 1 & 0.1 & 30 & 25 & 1 & Pro & $0^{\circ}$ \\
  3 & 0.1 & 1 & 0.1 & 50 & 34 & 1 & Pro & $0^{\circ}$ \\
  4 & 0.1 & 1 & 0.1 & 100 & 100 & 1 & Pro & $0^{\circ}$ \\
  5 & 0.2 & 1 & 0.1 & 50 & 36 &1 & Pro & $0^{\circ}$ \\
  6 & 0.1 & 1 & 0.1 & 50 & 40 &1 & Retro & $0^{\circ}$ \\
  7 & 0.1 & 1 & 0.1 & 30 & 30 & 1 & - & $90^{\circ}$ \\
  8 & 0.1 & 1 & 0.1 & 30 & 33 & 7 & Pro & $0^{\circ}$ \\
  9 & 0.1 & 1 & 0.5 & 40 & 10 & 1 & Pro & $0^{\circ}$ \\
 10 & 0.1 & 1.5 & 0.1 & 40 & 35 &1 & Pro & $0^{\circ}$ \\
  \hline
  \hline
\end{tabular}
\end{table*}

\subsection{SPH and the Hybrid Radiative Transfer Approximation}

\noindent Smoothed Particle Hydrodynamics (SPH) \citep{Lucy,Gingold_Monaghan,Monaghan_92} is a Lagrangian formalism that represents a fluid by a distribution of particles.  Each particle is assigned a mass, position, internal energy and velocity: state variables such as density and pressure can then be calculated by interpolation - see reviews by \citet{Monaghan_92,Monaghan_05}.  In these simulations, the gas is modelled by 500,000 SPH particles: the primary star (and the secondary companion) are represented by point mass particles, which can accrete the gas particles if sufficiently close to be bound \citep{Bate_code}.  

The SPH code used in this work is based on the SPH code developed by \citet{Bate_code}.  It uses individual particle timesteps, and individually variable smoothing lengths \(h_i\) such that the number of nearest neighbours for each particle is \(50 \pm 20\).  In the discs studied in this work, this provides a minimum disc scale height ($H$) to smoothing length ratio of $\frac{H}{h}\approx 1$ at the disc's inner edge (which increases to a maximum of approximately 6 at larger distances from the parent star).  The code uses a hybrid method of approximate radiative transfer \citep{intro_hybrid}, which is built on two pre-existing radiative algorithms: the polytropic cooling approximation devised by \citet{Stam_2007}, and flux-limited diffusion (e.g. \citealt{WB_1,Mayer_et_al_07}, see \citealt{intro_hybrid} for details).  This union allows the effects of both global cooling and radiative transport to be modelled without extra boundary conditions. 

The opacity and temperature of the gas is calculated using a non-trivial equation of state - this accounts for the effects of of H$_{2}$ dissociation, H$^{0}$ ionisation, He$^{0}$ and He$^{+}$ ionisation, ice evaporation, dust sublimation, molecular absorption, bound-free and free-free transitions and electron scattering \citep{Bell_and_Lin,Boley_hydrogen,Stam_2007}.  Heating of the disc can also be achieved by \(P\,dV\) work and by shocks.

\noindent For convenience, a brief recap of one of the test cases undertaken by the code is reproduced here .  This test case is the evolution of a protoplanetary disc, with parameters defined in a series of papers by \citet{Mejia_1,Mejia_2,Mejia_3,Mejia_4}.  The model is a \(0.07\,M_{\odot}\) Keplerian disc which extends to 40 AU, orbiting a star of \(0.5\,M_{\odot}\).  Initially, the surface density profile is \(\Sigma \sim r^{-1/2}\), with a temperature profile of \(T \sim r^{-1}\).  The disc is modelled using \(2.5 \times 10^5\) SPH particles, with one sink particle representing the star.  The disc is immersed in a radiation field of \(T_0(\mathbf{r}) = 3K\) ; the effects of disc irradiation by the central star are not included.

\begin{figure}
\begin{center}
\includegraphics[scale=0.39]{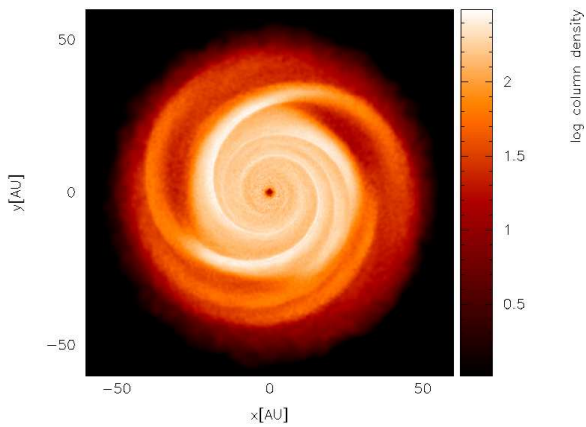}
\caption{Snapshot of the \citet{Mejia_3} test disc after 1906 years. \label{fig:boley}}
\end{center}
\end{figure}

A snapshot of the disc can be seen in Figure \ref{fig:boley}. The disc matches qualitatively the work of Boley et al, reproducing the burst and asymptotic phases, and producing a $Q$ profile consistent with their results.  A more detailed discussion can be found in \citet{intro_hybrid}.

\subsection{Initial Disc Conditions \label{sec:ICs}}

\noindent Originally, it was intended that the disc conditions used by \citet{Lodato_encounters} would be used to facilitate a direct comparison of techniques.  However, these initial conditions are scale-free, which poses problems when radiative transfer is simulated.  The \citet{Lodato_encounters} parameters were therefore modified to ensure that the disc is marginally gravitationally stable. In essence, this required decreasing the star mass, and increasing the radial extent of the disc.  The disc extends from \(r_{in} = 1\) au to \(r_{out} = 40 \) au.  The disc has mass \(0.1 \,M_{\odot}\), with a central primary star of mass \(0.5 \,M_{\odot}\).  The initial surface density profile was chosen to be \(\Sigma \propto r^{-1}\), with a sound speed profile of \(c_s \propto r^{-\frac{1}{4}}\).  Two variants of the disc were also run: one with a disc mass of \(0.2 \, M_{\odot}\), and one with  \(\Sigma \propto r^{-3/2}\).

\begin{figure*}
\begin{center}$
\begin{array}{cc}
\includegraphics[scale = 0.5]{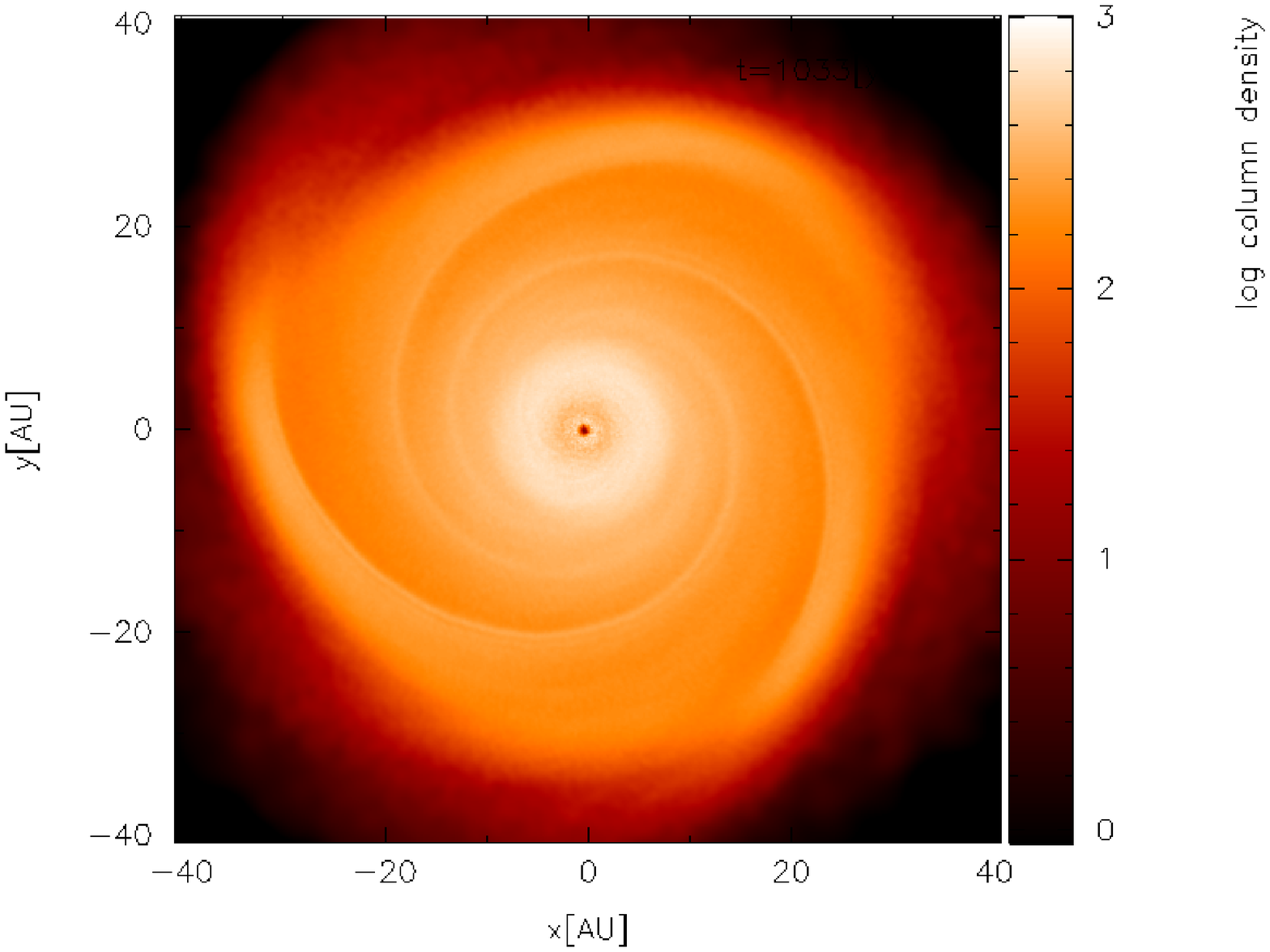} \\
\includegraphics[scale = 0.5]{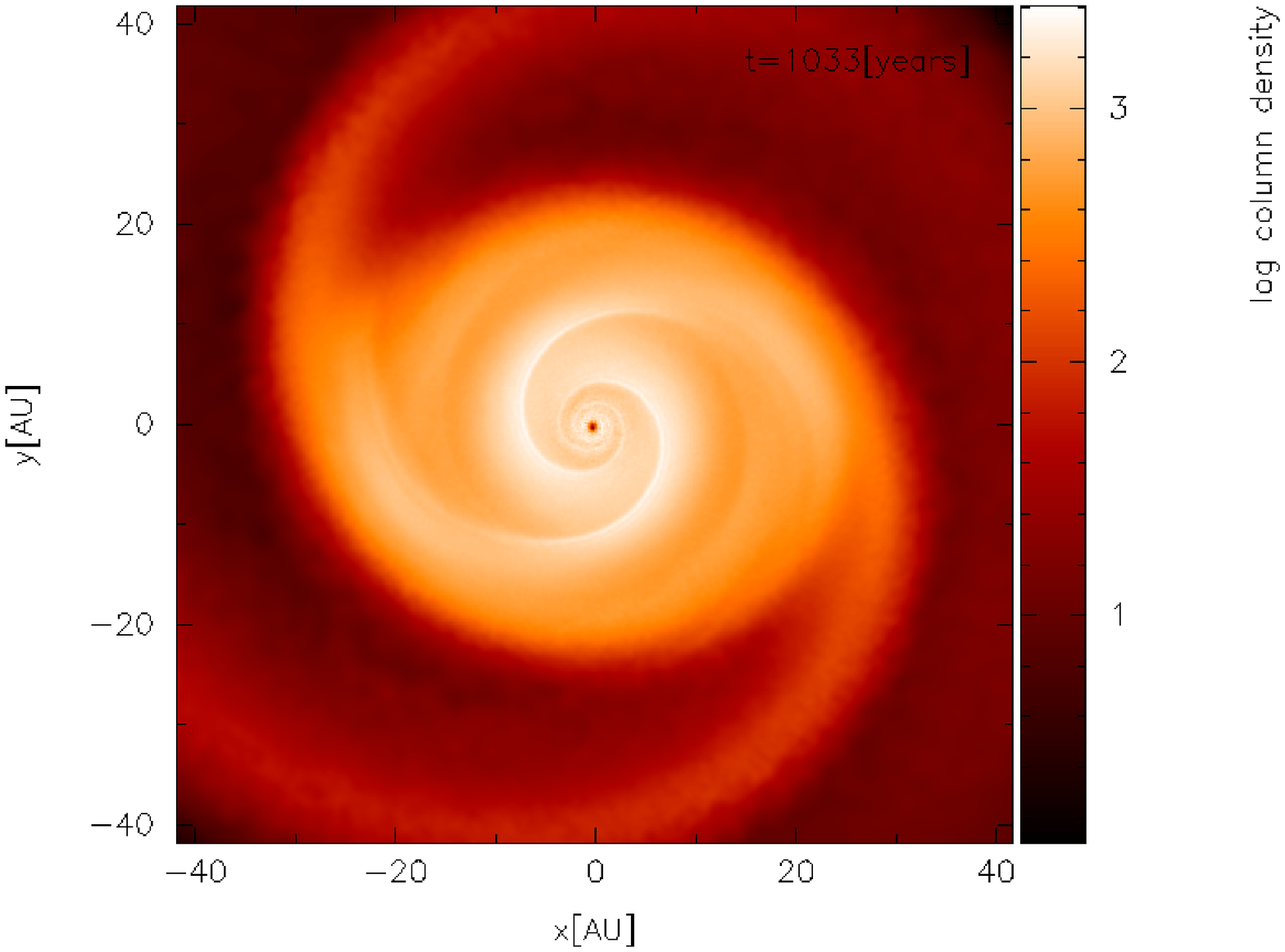} \\
\includegraphics[scale = 0.5]{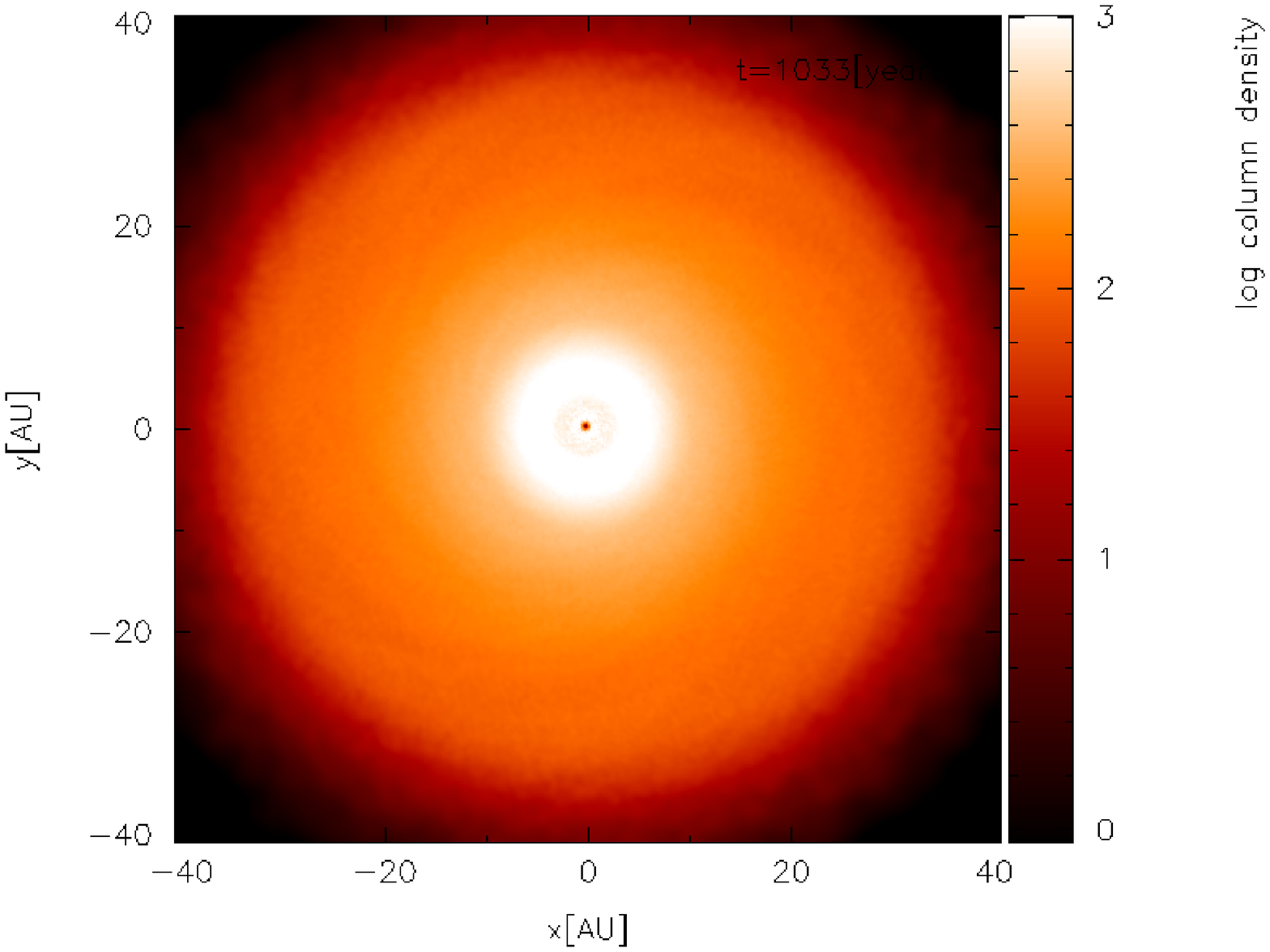} \\
\end{array}$
\caption{Snapshots of the three discs used after several ORPs: the \(0.1\, M_{\odot}\) disc (top), the \(0.2\, M_{\odot}\) disc (middle) and the \(\Sigma \propto r^{-3/2}\) disc (bottom).  Note each plot has a specific colour bar.\label{fig:disc}}
\end{center}
\end{figure*}

The discs were evolved in isolation for several Outer Rotation Periods (ORPs).  This allows the disc to approach an equilibrium state and become marginally unstable, and to develop steady-state spiral structures (with the exception of the  \(\Sigma \propto r^{-3/2}\) disc, see Figure \ref{fig:disc}).  As the discs are either stable or marginally unstable, the discs will require external stimuli in order to fragment.

\begin{figure*}
\begin{center}$
\begin{array}{cc}
\includegraphics[scale = 0.5]{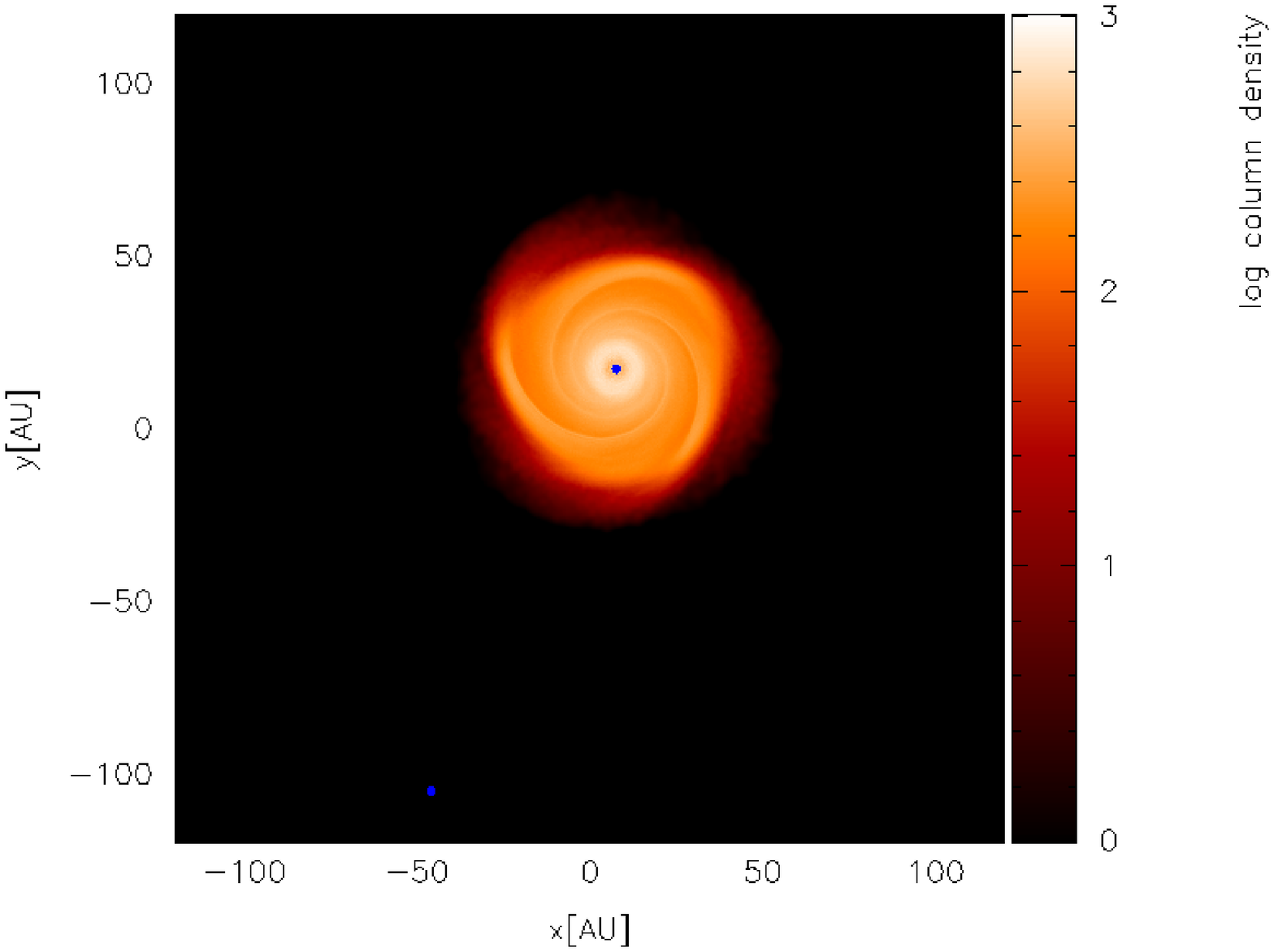} \\
\includegraphics[scale = 0.5]{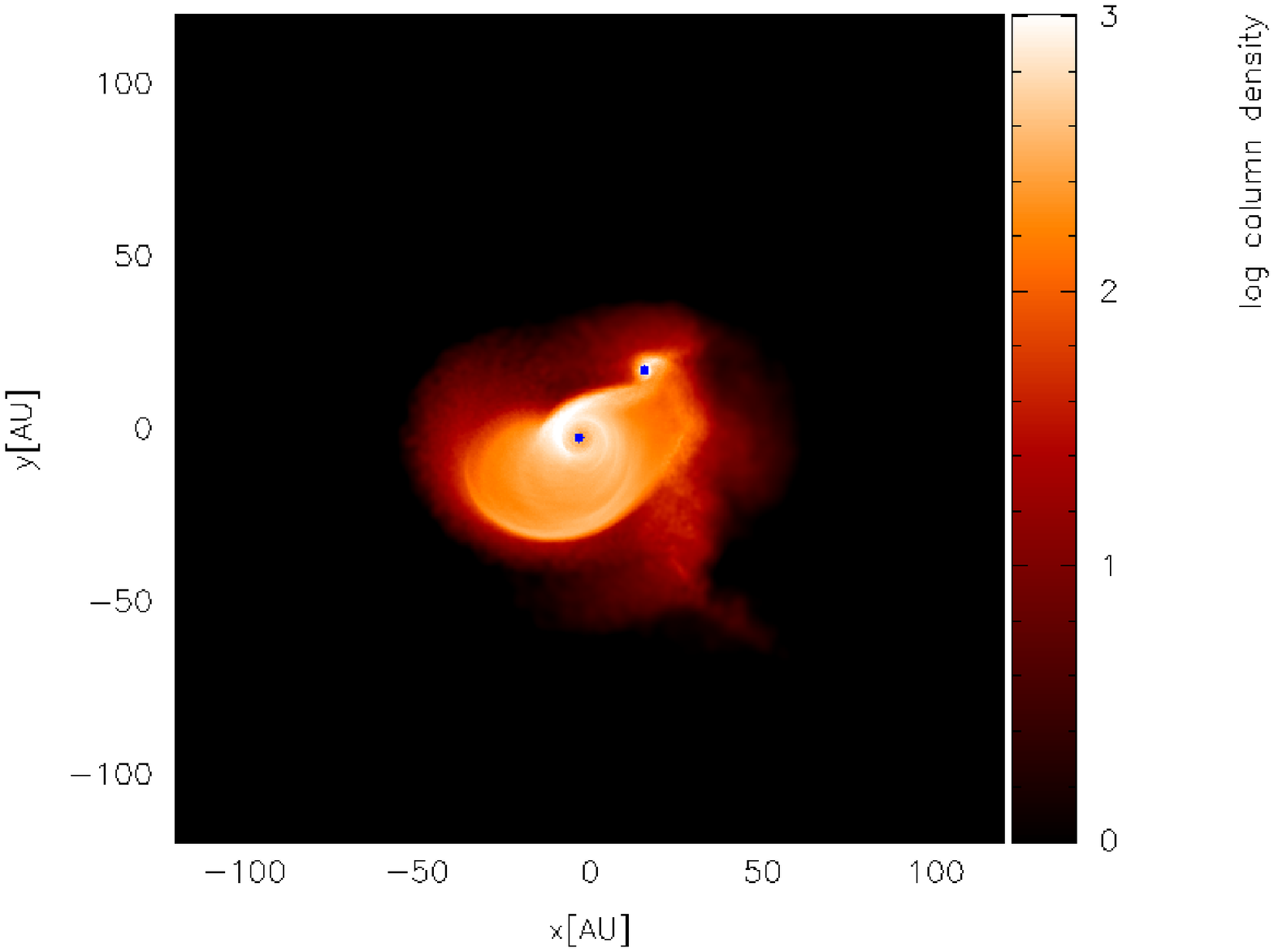} \\
\includegraphics[scale = 0.5]{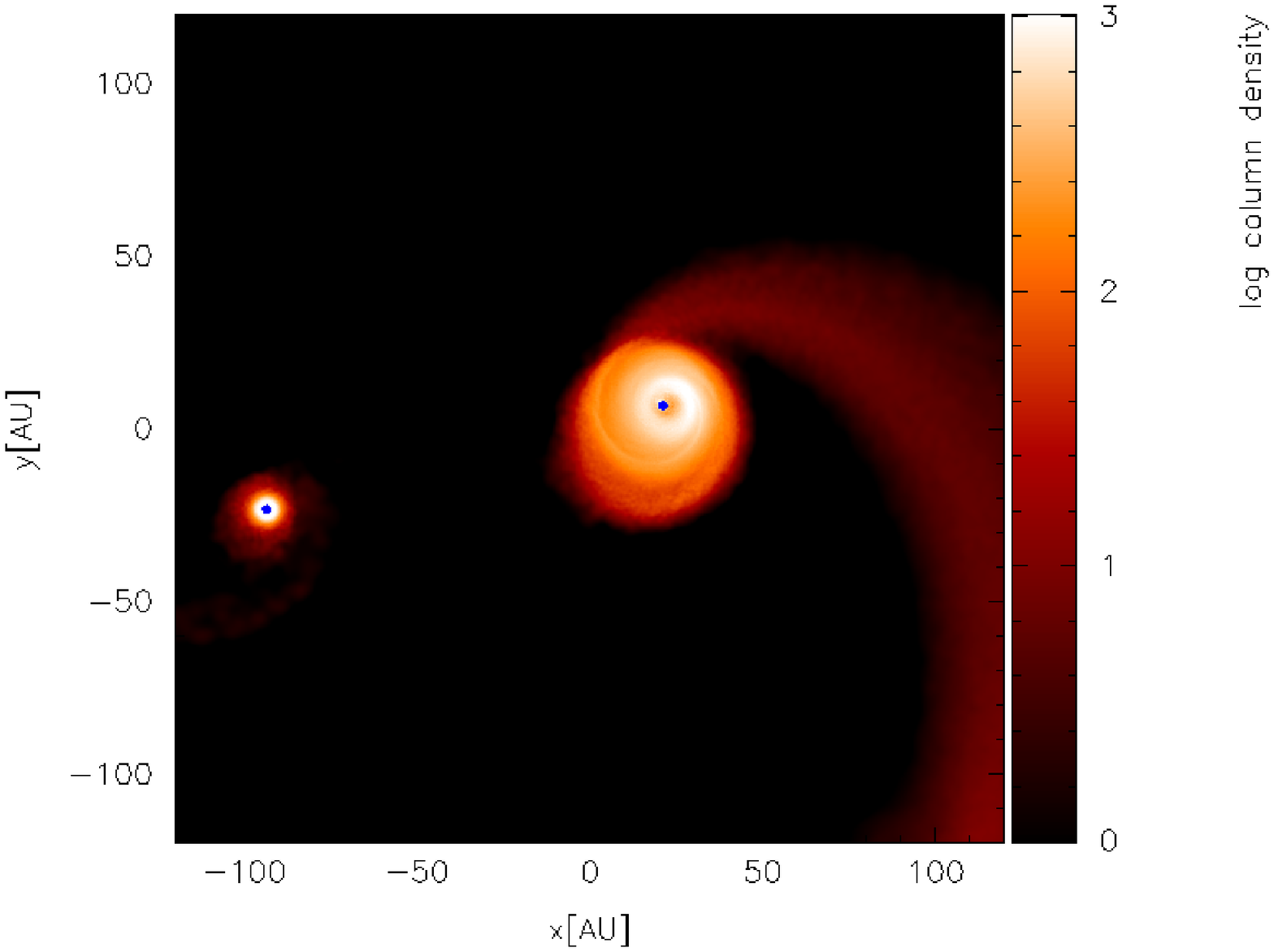} \\
\end{array}$
\caption{Images of the Simulation 1 disc before, during, and after the encounter (dot-dashed line). \label{fig:image_1}}
\end{center}
\end{figure*}	 

\subsection{The Stimulus: Adding a Companion}

\noindent With the disc evolved into a quasi-steady state, a secondary star was then added (at a separation of 100 au to the primary, in order to prevent any non-linear perturbations in the disc by the secondary's sudden appearance).  The secondary was added with a variety of orbital parameters, comprising a suite of 10 simulations (see Table \ref{tab:params} for details).  Of the encounters simulated, eight were coplanar, in prograde motion; one was coplanar, with retrograde motion; and one was inclined at an angle of \(90^{\circ}\).  Eight simulations were conducted using the standard \(0.1\) \(M_{\odot}\) disc, one was conducted using the \(0.2\) \(M_{\odot}\) disc, and one using the \(\Sigma \propto r^{-3/2}\) disc. This provides data on how disc perturbations vary with the disc mass, surface density profile, the secondary's periastron, eccentricity, inclination, and the relative angular momenta of the secondary and the disc.

\subsection{Resolution}

\noindent It is of utmost importance that any SPH simulations of disc fragmentation correctly resolve the fragments.  \citet{Burkert_Jeans} show that SPH correctly reproduces fragmentation if the minimum resolvable mass, 

\begin{equation} M_{min} = 2 N_{neigh}m_i = 2 M_{tot} \frac{N_{neigh}}{N_{tot}} \end{equation}

\noindent is less than the Jeans mass of interest (\(m_i\) is the mass of a single SPH particle, \(N_{tot}\) is the total number of particles, \(N_{neigh}\) is the typical number of nearest neighbours for a single SPH particle, and \(M_{tot}\) is the total mass).  The Jeans mass can be estimated using the fact that the most unstable wavelength to gravitational instability is the disc's scale height, \(H\).  This gives a Jeans Mass \(M_{J} = \Sigma H^2 = M_{tot}\left(\frac{H}{R}\right)^2\), and hence

\begin{equation} \frac{M_{J}}{M_{min}} \approx \frac{1}{2} \left(\frac{H}{R}\right)^2 \frac{N_{tot}}{N_{neigh}} \approx \frac{1}{2} \left(\frac{M_{disc}}{M_{*}}\right)^2 \frac{N_{tot}}{N_{neigh}} \end{equation}

\noindent All the simulations use 500,000 particles, and have \(N_{neigh} = 50\).  Using \(M_{disc} = 0.1 M_{\odot}\), and \(M_{*} = 0.5 M_{\odot}\), this gives \(M_{J} \approx 200 M_{min}\).  With the Jeans mass of interest being two orders of magnitude above the minimum resolved mass, this shows that all simulations carried out in this work easily satisfy the necessary resolution conditions.

\section{Results}\label{sec:results}

\noindent The results for each simulation are discussed below.  A summary of all simulations can be found in Table \ref{tab:params}.  Qualitatively speaking, the results show features common to all simulations.  Therefore, the results for simulation 1 (the reference simulation) will be discussed in detail - the other simulations will be more briefly described, focusing on their unique and differing features.

\subsection{Simulation 1 - The Reference Simulation}

\begin{figure}
\includegraphics[scale=0.5]{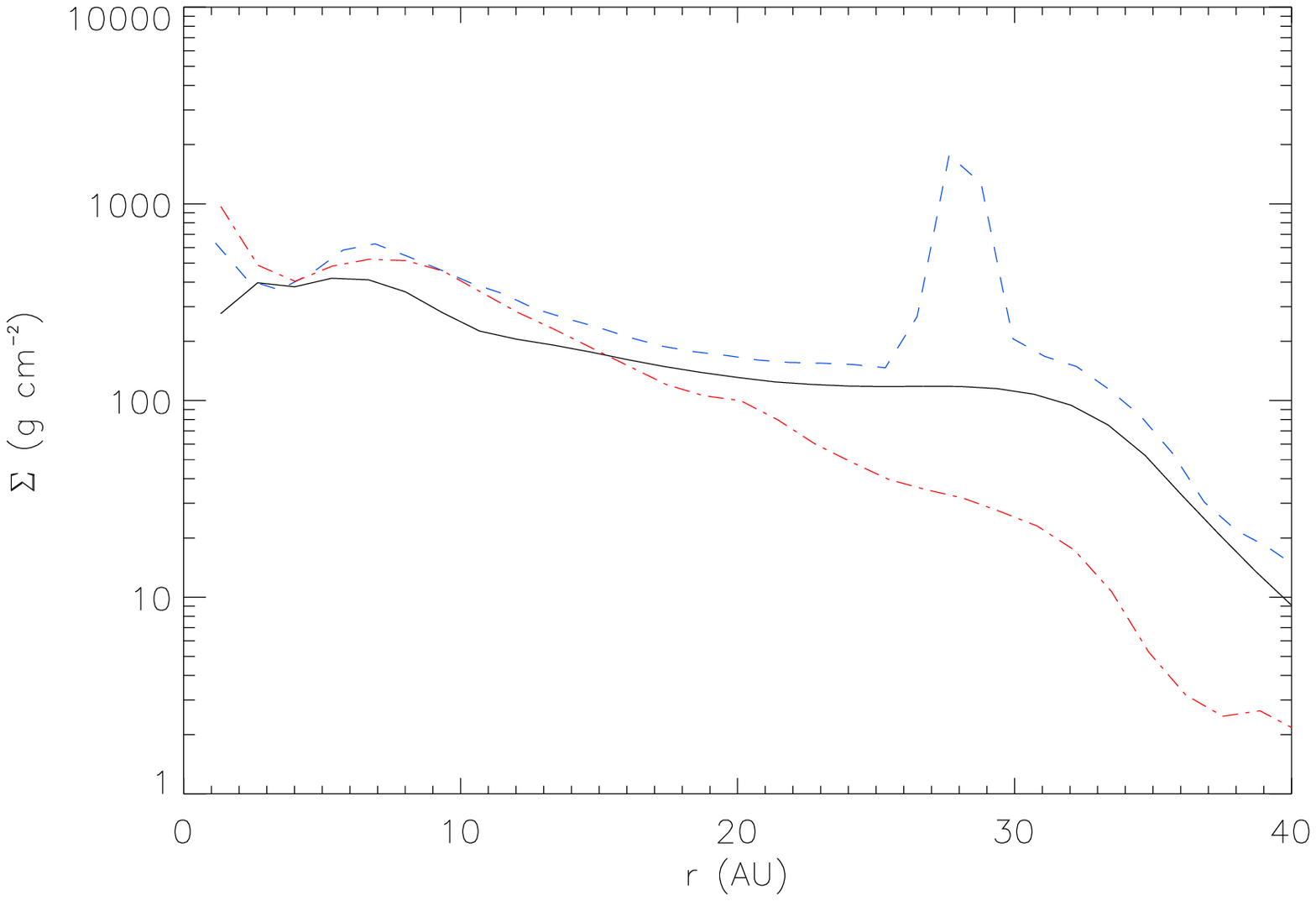}
\caption{Surface density profile of the Simulation 1 disc before the encounter (solid line), at periastron (dashed line), and after the encounter (dot-dashed line). \label{fig:sim_1sigma}}
\end{figure}

\begin{figure}
\includegraphics[scale=0.5]{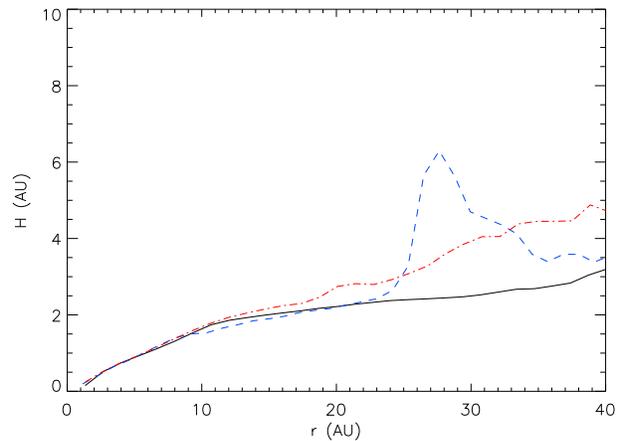}
\caption{Scale height of the Simulation 1 disc before the encounter (solid line), at periastron (dashed line), and after the encounter (dot-dashed line). \label{fig:sim_1scale}}
\end{figure}

\begin{figure}
\includegraphics[scale=0.5]{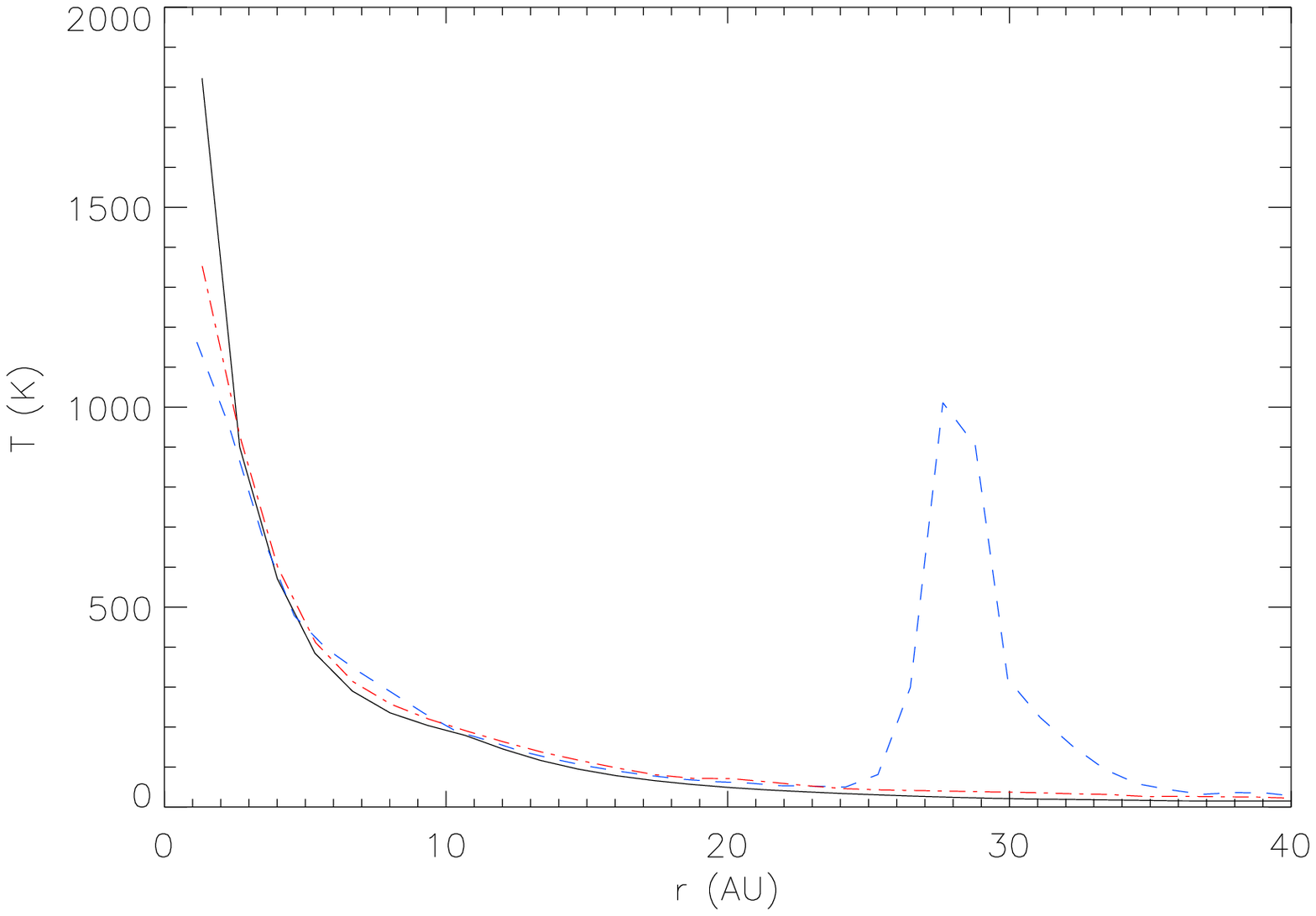}
\caption{Midplane temperature profile of the Simulation 1 disc before the encounter (solid line), at periastron (dashed line), and after the encounter (dot-dashed line). \label{fig:sim_1T}}
\end{figure}

\begin{figure}
\includegraphics[scale=0.5]{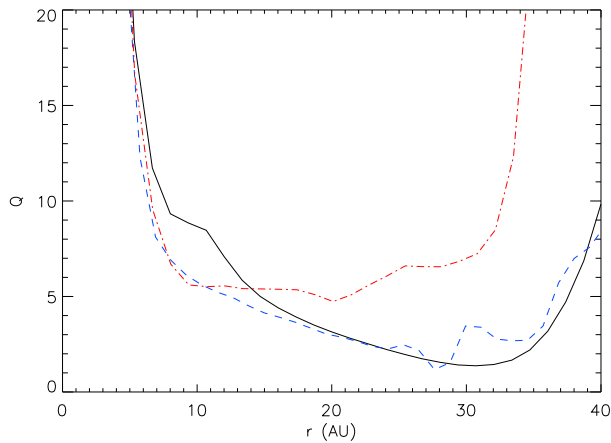}
\caption{Toomre \(Q\) profile of the Simulation 1 disc before the encounter (solid line), at periastron (dashed line), and after the encounter (dot-dashed line). \label{fig:sim_1Q}}
\end{figure}

\noindent Images of the reference simulation can be seen in Figure \ref{fig:image_1}: the azimuthally-averaged radial profiles at the three instants are shown in Figures \ref{fig:sim_1sigma}, \ref{fig:sim_1scale}, \ref{fig:sim_1T}, \& \ref{fig:sim_1Q}.  As the secondary passes through the disc, it imparts significant energy to the disc, causing a strong temperature spike at the location of the secondary, and the scale height of the disc to be enlarged.  The disc exerts significant tidal forces on the secondary as it reaches periastron, and angular momentum is transferred to the disc.  There are three consequences to this: 

\begin{enumerate}
\item The orbital parameters are altered, such that although the initial velocity of the secondary is consistent with \(r_{peri} = 40\) au, the actual periastron is lower (this can be seen in the \(\Sigma\) spike in Figure \ref{fig:sim_1sigma}). 
\item The secondary can capture disc matter to form a secondary disc (\(M\sim 0.006 M_{\odot}\), comprising around 5\% of the initial disc), as well as drawing out tidal tails (see Figure \ref{fig:image_1}), which depletes the primary disc.  In total, around 20\% of the initial disc mass is lost.
\item The disc is pushed out of equilibrium, and angular momentum transport is required to stabilise the disc: this results in a steeper surface density profile (and a flatter \(Q\) profile).  This is a common feature to all the simulations carried out (although the strength of these effects varies with the orbital parameters used).
\end{enumerate}

\begin{figure}
\includegraphics[scale=0.5]{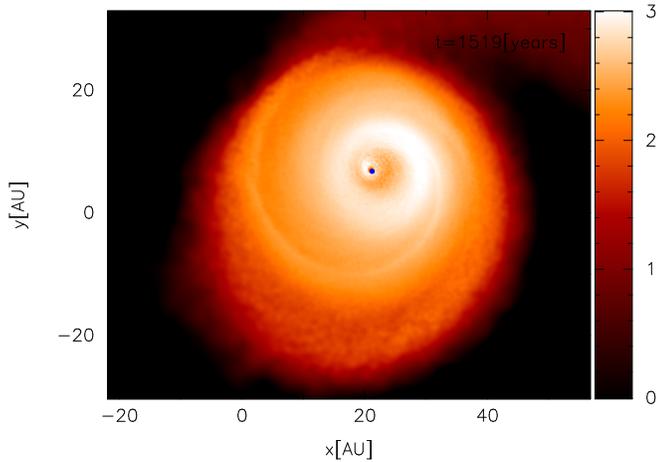}
\caption{The Simulation 1 disc after the encounter. \label{fig:sim_1after}}
\end{figure}

\noindent Once the encounter is complete, the disc spiral structure is much weaker (Figure \ref{fig:sim_1after}), and the disc has stabilised at all radii, with a higher surface density and scale height, and an unchanged temperature profile.

\subsection{Simulation 2 - A Low Periastron Encounter}

\noindent With a low periastron radius in prograde motion, the disc exerts stronger tidal forces, resulting in the capture of the secondary on a highly eccentric orbit.  The secondary captures a similar disc to Simulation 1, but the total disc mass lost is slightly higher (due to more massive tidal tails being induced).  This increased loss results in a larger steepening of the surface density profile, as the primary disc must make a larger readjustment to retain stability.

\subsection{Simulation 3 - A High Periastron Encounter}

\noindent A high periastron encounter, although having the same qualitative features as Simulations 1 \& 2, has a reduced quantitative effect.  This is of course consistent with the gravitational influence of an object decreasing with separation.  The surface density profile is steepened, but the effect is only noticeable at large radii, which gives a \(Q\) profile that remains flat out to around 35 AU (slightly further than Simulation 1).

\subsection{Simulation 4 - A Distant Periastron Encounter}

\begin{figure}
\includegraphics[scale=0.5]{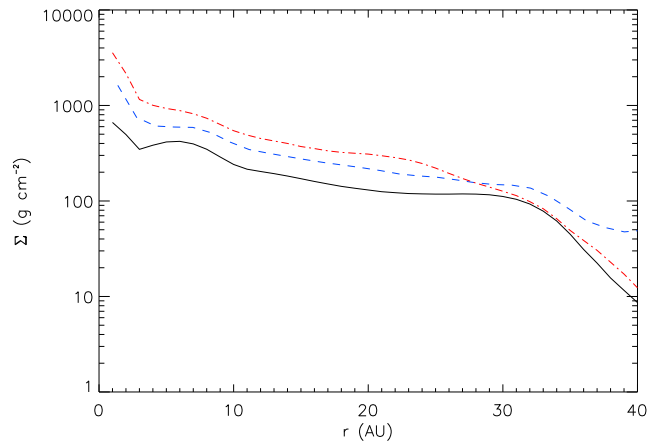}
\caption{Surface density profile of the Simulation 4 disc before the encounter (solid line), at periastron (dashed line), and after the encounter (dot-dashed line). \label{fig:sim_4sigma}}
\end{figure}

\begin{figure}
\includegraphics[scale=0.5]{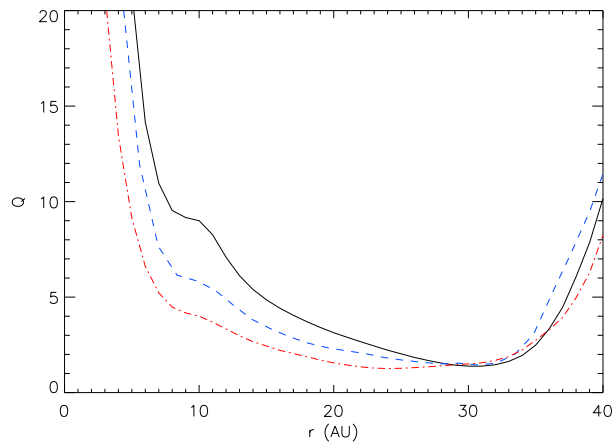}
\caption{Toomre \(Q\) profile of the Simulation 4 disc before the encounter (solid line), at periastron (dashed line), and after the encounter (dot-dashed line). \label{fig:sim_4Q}}
\end{figure}

\noindent When the periastron is very large, the effects of the encounter are reduced significantly.  However, this simulation is worth discussing. The distance of the secondary from the disc ensures that the disc heating is minimal, which allows the outer regions of the disc to remain marginally unstable.  Also, the surface density profile is slightly modified in the inner regions (Figure \ref{fig:sim_4sigma}), without affecting the scale height or the temperature profile.  This results in a disc that is marginally unstable over a larger range of radii (Figure \ref{fig:sim_4Q}).  Although the disc does not fragment in this simulation, it has been modified by the encounter to become more amenable to fragmentation.  This suggests that there is a range of periastra for which encounters are most effective at destabilising the disc, which will depend on the disc's (and the secondary's) properties. 

\subsection{Simulation 5 - A Higher Disc Mass Encounter}

\begin{figure}
\includegraphics[scale=0.5]{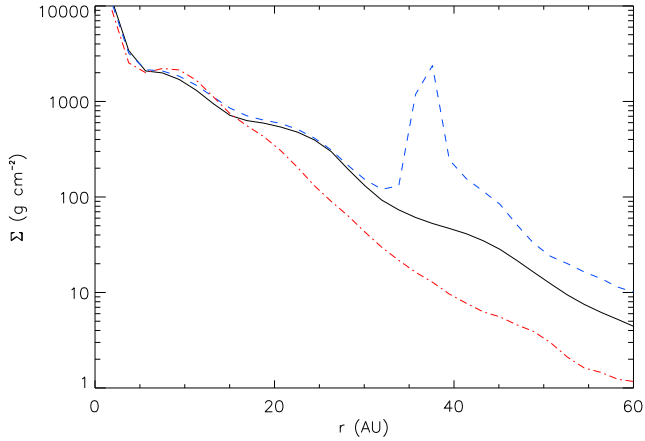}
\caption{Surface density profile of the Simulation 5 disc before the encounter (solid line), at periastron (dashed line), and after the encounter (dot-dashed line). \label{fig:sim_5sigma}}
\end{figure}

\begin{figure}
\includegraphics[scale=0.5]{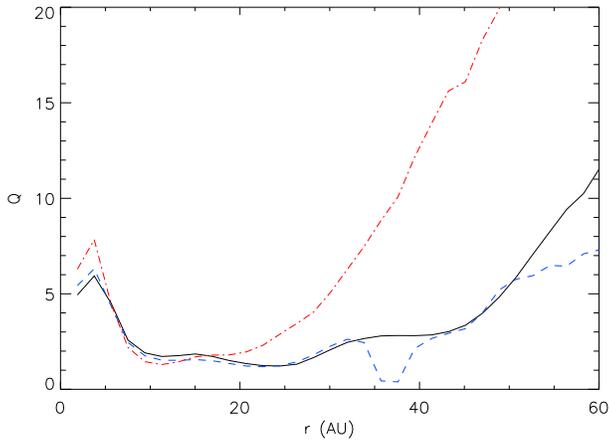}
\caption{Toomre \(Q\) profile of the Simulation 5 disc before the encounter (solid line), at periastron (dashed line), and after the encounter (dot-dashed line). \label{fig:sim_5Q}}
\end{figure}

\noindent The previous simulations have shown that the effect of an encounter is to steepen the disc's surface density profile, and to flatten the \(Q\) profile.  Will this behaviour hold for higher disc masses? To investigate this, a prograde coplanar encounter was simulated using the \(0.2\) \(M_{\odot}\) disc described in section \ref{sec:ICs}.  Figure \ref{fig:sim_5sigma} shows that the surface density profile does steepen, but not a great deal: this is primarily due to significant mass depletion (and relatively reduced mass redistribution) - the secondary captures a disc of \(0.01\,M_{\odot}\), removing matter from the outer regions to steepen the profile.  At periastron, a significant increase of the surface density (and scale height) can be seen; the increased density at periastron (in comparison with Simulation 1) prevents the temperature increase from reaching the same magnitude.  The \(Q\) profile for the disc (Figure \ref{fig:sim_5Q}) was initially rather flat; during the encounter, the companion induces a small region to become very unstable around periastron, but due to the mass stripping, the majority of the disc stabilises: the region of instability decreases from around \(10 - 40 \) au to around \(10 - 20\) au.  

\subsection{Simulation 6 - A Retrograde Encounter}

\begin{figure}
\includegraphics[scale=0.5]{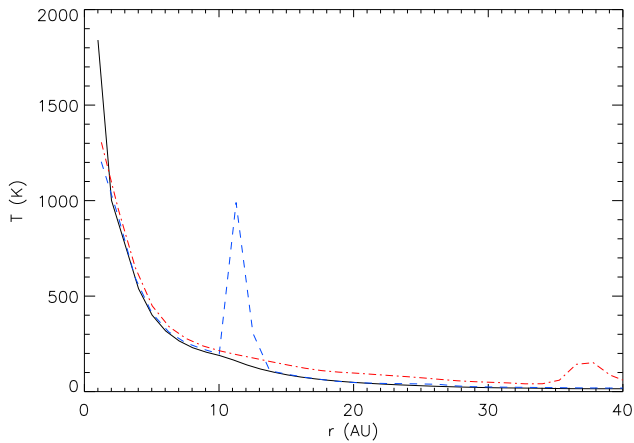}
\caption{Midplane temperature profile of the Simulation 6 disc before the encounter (solid line), at periastron (dashed line), and after the encounter (dot-dashed line). \label{fig:sim_6T}}
\end{figure}

\noindent In prograde encounters, the ability of the disc spiral structure to couple with the perturber is much improved in comparison with the retrograde encounter, and hence angular momentum transfer between the perturber and the disc in a retrograde encounter is relatively smaller \citep{Hall_96}.  In general, this is reflected in the results obtained.  However, depending on the relative phases of the spiral density wave and the secondary, the secondary can encourage the spirals to wind more tightly, allowing compressive heating in the inner regions.  This is evident in the temperature spike at \(\sim\) 10 au in Figure \ref{fig:sim_6T}.  Despite this, the inner regions of the disc appear to be less stable than they were initally.  This dependence on the relative orbital phases shows that the influence of encounters on disc dynamics are indeed more complex than a simple study of orbital parameters (such as this work) may initially suggest. 

\subsection{Simulation 7 - A High Inclination Encounter}

\noindent As a secondary on a high inclination orbit has a short disc interaction timescale, it should be expected that the strength of the perturbation is reduced in comparison to a coplanar orbit.  Our results do seem to confirm this.  Although the outer regions of the disc appear to increase in mass (and become less stable), the inner regions of the disc are almost unaffected: the \(Q\) and \(\Sigma\) profiles remain similar in the region \(10-30\) au, while the temperature profile is unaffected at virtually all radii.  It could be argued that the transport of mass outward out of optically thick regions (allowing more efficient cooling) is favourable to fragmentation, however none was seen.

\subsection{Simulation 8 - A Hyperbolic Encounter}

\begin{figure}
\includegraphics[scale=0.5]{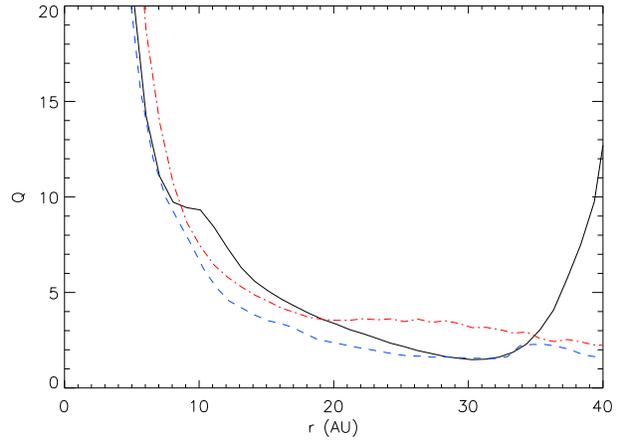}
\caption{Toomre \(Q\) profile of the Simulation 8 disc before the encounter (solid line), at periastron (dashed line), and after the encounter (dot-dashed line). \label{fig:sim_8Q}}
\end{figure}

\noindent As with the high inclination encounter, a hyperbolic encounter has a shorter interaction timescale than a parabolic encounter, and so should exert less influence on the disc dynamics.  Figure \ref{fig:sim_8Q} does indeed show that the perturbation to Q is weak in comparison with Simulation 1: however, the impact of the high-velocity encounter has resulted in significant mass stripping, causing a decreased Q in the outer regions: however, the velocity dispersion of the outflow ensures that the gas is still unbound. 

\subsection{Simulation 9 - A High Mass Secondary Encounter}

\begin{figure}
\includegraphics[scale=0.5]{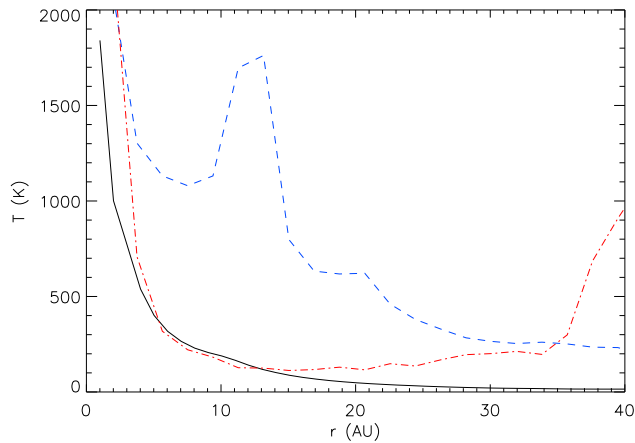}
\caption{Midplane temperature profile of the Simulation 9 disc before the encounter (solid line), at periastron (dashed line), and after the encounter (dot-dashed line). \label{fig:sim_9T}}
\end{figure}

\noindent For the last of the simulations that emulate \citet{Lodato_encounters}, the effect of increasing the secondary mass is studied here.  With the primary and secondary now of equal mass, the disc begins to feel the influence of the encounter much earlier than in previous simulations.  A substantial tidal tail is formed at an early time, and becomes significantly concentrated.  This provides significant torques at much earlier times than the other simulations, which results in a far smaller periastron (\(\sim 10\) au), and at greater speed, which in turn heats the inner disc significantly. (see Figure \ref{fig:sim_9T}).  Very efficient mass-stripping causes the disc to decrease in mass by more than half: a large fraction has been swept into the tidal tail at larger radii, and into a secondary disc.  The significant mass loss and heating ensures the primary disc itself cannot fragment - it is now only a few au in radius, and extremely hot.

This simulation would suggest that high mass encounters are even less likely to promote fragmentation than low mass encounters (at least with periastra within the disc).  It is possible, however, that a distant encounter with a high mass secondary would be more successful at producing fragments (cf. Simulation 4), although with a high mass secondary's larger perturbative potential, the periastron would most likely need to be larger than is considered here, and an interesting avenue for further work. 

\subsection{Simulation 10 - A Steep Disc Profile Encounter}

\noindent Having studied the results of the previous simulations, it seems that the result of most encounters is to steepen the surface density profile of the disc.  This behaviour then begs the question: what if the profile is steeper initially? Does the encounter result in an even steeper profile, or does there exist an asymptotic profile to which a disc will tend (given enough encounters)?  To answer this question, an additional simulation was run with a disc exhibiting a \(\Sigma \propto r^{-3/2}\) profile.  The disc was run in isolation (as with the other discs used in this work) for the same timescale: this does mean however that the disc is stable over most radii when the encounter is begun (although \(Q\) is tending towards instability at larger radii).  Again, this simulation was run for the purpose of studying the effects of encounters on the surface density profile, not fragmentation, so the need for a marginally stable disc is not so important here. 

\begin{figure}
\includegraphics[scale=0.5]{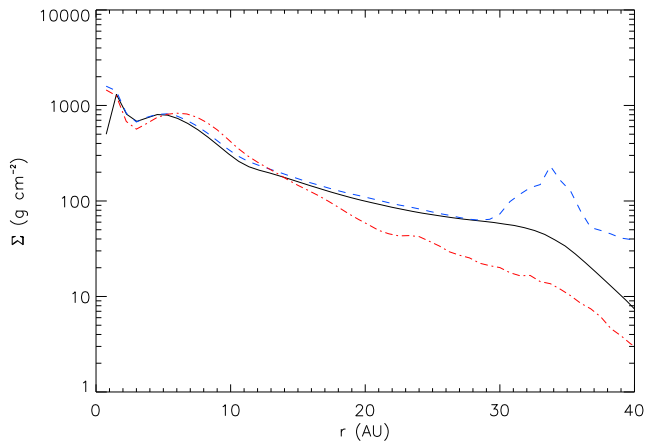}
\caption{Surface density profile of the Simulation 10 disc before the encounter (solid line), at periastron (dashed line), and after the encounter (dot-dashed line). \label{fig:sim_10sigma}}
\end{figure}

As can be seen in Figure \ref{fig:sim_10sigma}, the effect of the encounter still steepens the profile (although comparing with Simulation 1 (Figure \ref{fig:sim_1sigma}, the magnitude of disc readjustment is indeed smaller).  As the outer regions of the disc are lower in density than in Simulation 1, the scale height is more sensitive to the encounter, and therefore see a greater increase as the secondary passes through the disc.  However, the low density also ensures that compressive heating is less effective: therefore a lower temperature increase is seen. 

This additional simulation has shown that surface density steepening appears to be common to all discs, regardless of their initial profile.  However, the steeper the disc initially, the less effect the encounter has: this does suggest that there may be a profile steep enough that an encounter does not affect it significantly - but, this putative profile may be too steep to be physically appropriate.

\section{Discussion}\label{sec:discussion}

\noindent Having displayed results from a number of simulations, the broad trends uncovered by this work will now be discussed.

\subsection{The Influence of Disc Mass, Disc Profile and Secondary Mass}

\noindent Changing the mass involved in the encounter (and its distribution in the system) will of course affect the tidal forces experienced by all participants in the encounter.  Increasing the disc mass might increase its initial instability, but this does not in general indicate a more unstable disc after an encounter.  The inner regions of a more massive disc can remain more gravitationally unstable, but the inefficiency of cooling in these regions precludes the possibility of fragmentation \citep{Gammie,Ken_1,Rafikov_05,Rice_and_Armitage_09}.

Encounters in general steepen the surface density profile: this appears to be true regardless of the initial profile, but the magnitude of profile steepening does seem to decrease if the initial profile is steep.  They also flatten the \(Q\) profile as a secondary effect, but this does not appear to be true if the Q profile is initially flat.

As far as secondary mass is concerned, the results shown here show that its primary influence is increased tidal forces subjected on the disc (as well as increased compressive heating at periastron).  Therefore, for disc-penetrating encounters, a larger secondary has a deleterious effect on potential fragmentation in comparison to a smaller secondary.  

\subsection{The Influence of Periastron Radius}

\noindent A crude expectation may be that reducing periastron radius will destabilise the disc more, and induce fragmentation more readily.  This is incorrect, primarily due to radiative effects.  The ability of the disc to cool has been shown to be an important factor.  If the periastron is too low, the disc is heated too strongly by the secondary's motion through it, and the disc is readjusted into a much more stable state due to mass stripping and angular momentum redistribution.  If the periastron is too high, the disc will not feel the effects of the secondary's motion, and the disc will feel no stimulus at all.  However, if the periastron is within some range of values (determined by the properties of the primary, its disc and the secondary), then the result of the encounter is to \emph{increase} the region of instability in the disc.  An interesting question that can now be asked is: would repeated encounters lead to fragmentation? The answer is beyond the scope of this paper, but the results would indicate that repeated encounters would in general increase outward angular momentum transport (and inward mass transport), pushing the surface density profile to an equilibrium value that is too steep for gravitational instability to act, as the available mass at large radii is insufficient \citep{Clarke_09,Rice_and_Armitage_09}.  This hypothesis requires testing, and is an interesting avenue for further work.

\subsection{The Influence of Angular Momentum Alignment (and Inclination)}

\noindent As has been previously discovered \citep{Hall_96,Lodato_encounters}, the most effective encounters are prograde in nature, and coplanar to the disc.  However, the results of this work show that retrograde encounters may be more complex than initially thought.  The effect of spiral winding by the secondary can encourage compressive heating in the inner regions, and the angular momentum transport in this case allows a disc that is more unstable in the inner regions.  However, there is no indication that fragmentation would be better achieved using anything other than prograde encounters. 

Encounters outside of the disc plane (Simulation 7) can drag matter to larger vertical distance, increasing the disc scale height: but, the interaction time is insufficient for the encounter to encourage significant change.

\subsection{The Possibility of Binary Capture}

\begin{table*}
\centering
  \caption{Modification of the dynamical properties as a result of the encounter, for a selection of simulations.\label{tab:capture}}
  \begin{tabular}{c || ccccc}
  \hline
  \hline
   Simulation  &  $E_{i}$ (code units) & $E_{f}$ (code units)  &   $\Delta e$   & Secondary Disc? &   Tidal Tail?\\  
 \hline
  2 (30 AU) & $-3.2\times 10^{-4}$ & $-6.7\times 10^{-4}$ & -0.225 & Yes & Yes \\
  1 (40 AU) & $-2.4\times 10^{-4}$ & $-4.6\times 10^{-4}$ & -0.199 & Yes & Yes \\
  3 (50 AU) & $-1.9\times 10^{-4}$ & $-3.7\times 10^{-4}$ & -0.183 & Yes & Yes \\
  4 (100 AU) & $-9.5\times 10^{-5}$ & $-1.2\times 10^{-4}$ & -0.009 & Negligible & Negligible \\ 
  6 (Retro) & $-4.5\times 10^{-4}$ & $-8.5\times 10^{-4}$ & -0.175 & Negligible & Yes  \\
  8 (Hyperbolic) & $ 2.4\times 10^{-3}$ & $2.2\times 10^{-3}$ & +0.16 & No & Yes \\
  \hline
  \hline
\end{tabular}
\end{table*}

\noindent It is difficult to make an encounter precisely parabolic for numerical reasons.  With the exception of Simulation 8, the encounters simulated in this work have all been almost parabolic: that is, the total energy of the secondary is close to zero, and that any decrease in this total energy will result in the secondary becoming more bound to the primary-disc system.  The efficiency of binary capture can be estimated from the change in total energy experienced as a result of the encounter, as well as the modification of the orbital eccentricity\footnote{These eccentricities are calculated at the initial and final time of each simulation assuming two bodies in the system: i) the primary and its disc, and ii) the secondary and its disc (if any) - the tidal tail is not considered.}.  Table \ref{tab:capture} shows the initial total energy of the secondary ($E_{i}$), the total energy of the secondary after the encounter ($E_{f}$), and the change in orbital eccentricity ($\Delta e$).  The total energy decreases by a factor of 2 in most cases.  This appears to be due to the angular momentum transfer between the secondary and the disc it accretes.  As the secondary leaves the primary-disc system, in order to extract material orbiting in the same locale (i.e. to allow it to achieve escape velocity) it must impart dynamical energy to the gas.  Also, the process of exciting a tidal tail (from matter at the opposite orbital phase) requires angular momentum from the secondary, although this appears to be less important.  Encounters that do not create a sufficient disc (e.g. distant or high-velocity encounters) do not experience a significant orbital modification.

These simulations show that capture efficiency increases with decreasing distance, provided that the velocity of the orbit is low enough that the secondary captures a non-negligible disc of its own.  If the orbit is initially strongly unbound (such as Simulation 8), then no secondary disc can be captured, and the binary formation efficiency is effectively zero.  Furthermore, a high velocity encounter causes severe disc stripping, which diminishes the possibility of a future encounter resulting in capture.  These results are consistent with the work of \citet{Clarke_Pringle_binary}, who study the upper limit of capture efficiency for hyperbolic orbits, and find that the capture rate in those cases is indeed low.

\section{Conclusions}\label{sec:conclusions}

\noindent This paper has detailed the results of Smoothed Particle Hydrodynamics (SPH) simulations with hybrid radiative transfer \citep{intro_hybrid} of the encounters between a protostellar disc system and a secondary star.  The influence of different disc parameters on the possibility of disc fragmentation was explored, and the data is summarised above.  In general: \emph{encounters do not induce fragmentation} (confirming the results of \citealt{Lodato_encounters}), although there appears to be a subset of orbital parameters that modify the disc to make it more unstable over a larger range of radii.  As to whether this is an indication that there are specific orbital parameters for an encounter that cause fragmentation (e.g. a highly eccentric elliptical coplanar orbit which has a distant non-penetrating periastron), more work needs to be done.  Equally, the results presented here show that factors not incorporated in this parameter study (such as the relative orbital phases of the primary, the disc spiral structure and the secondary) have a non-negligible role to play in the resulting disc dynamics, indicating more work is required to fully understand their effects.  It has been shown that angular momentum transfer between the secondary and the disc is significant: however, calculations indicate that only around 1 in 10,000 discs will experience encounters of this nature in 1 Myr \citep{Clarke_Pringle_binary}, so this cannot be a typical trigger for angular momentum transport in discs.

The most important conclusion to draw from this work is the same as was found by \citet{Lodato_encounters} - the key parameter that determines disc fragmentation is the cooling rate \citep{Gammie,Ken_1,Mejia_2,Rice_et_al_05}.  This is independent of whether discs become unstable in isolation, or are driven to it by external stimuli such as stellar encounters.

\section*{Acknowledgements}

\noindent All simulations presented in this work were carried out using high performance computing funded by the Scottish Universities Physics Alliance (SUPA).  Surface density plots were made using \begin{small}{SPLASH}\end{small} \citep{SPLASH}.  The authors would like to thank the anonymous referee for their useful comments which strengthened this paper.

\bibliographystyle{mn2e} 
\bibliography{encounters}

\label{lastpage}

\end{document}